\documentclass[conference]{IEEEtran}

\usepackage{amsmath}

\usepackage{amsfonts}

\usepackage{amssymb}

\usepackage{stmaryrd}

\usepackage[mathscr]{eucal}

\usepackage{amsbsy}

\usepackage{bm}

\usepackage{bbding}

\usepackage{array}

\usepackage{tabularx}

\usepackage{ctable}

\usepackage[neveradjust]{paralist}

\usepackage{fancyvrb}

\usepackage{caption}
\usepackage{subfig}

\usepackage{placeins}

\usepackage{ifpdf}
\ifpdf
\DeclareGraphicsExtensions{.pdf,.png}
\else
\DeclareGraphicsExtensions{.eps}
\fi

\usepackage{algpseudocode}

\newcommand{\Tra}{^{{\sf T}}} 
\newcommand{\V}[1]{{\bm{\mathbf{\MakeLowercase{#1}}}}} 
\newcommand{\VE}[2]{\MakeLowercase{#1}_{#2}} 

\newcommand{\M}[1]{{\bm{\mathbf{\MakeUppercase{#1}}}}} 
\newcommand{\MC}[2]{\V{#1}_{#2}} 

\newcommand{\T}[1]{\boldsymbol{\mathscr{\MakeUppercase{#1}}}} 

\newcommand{\KOp}[1]{\llbracket #1 \rrbracket} 

\newcommand{\norm}[1]{\left\lVert \, #1 \, \right\rVert}

\newcommand{\email}[1]{\href{mailto:#1}{\nolinkurl{#1}}}

\newcommand{\Sec}[1]{\hyperref[sec:#1]{\S\ref*{sec:#1}}} 
\newcommand{\Section}[1]{\hyperref[sec:#1]{Section~\ref*{sec:#1}}} 
\newcommand{\AppFull}[1]{\hyperref[sec:#1]{Appendix~\ref*{sec:#1}}} 
\newcommand{\Eqn}[1]{\hyperref[eq:#1]{(\ref*{eq:#1})}} 
\newcommand{\Fig}[1]{\hyperref[fig:#1]{Figure~\ref*{fig:#1}}} 
\newcommand{\Tab}[1]{\hyperref[tab:#1]{Table~\ref*{tab:#1}}} 
\newcommand{\Thm}[1]{\hyperref[thm:#1]{Theorem~\ref*{thm:#1}}} 
\newcommand{\Cor}[1]{\hyperref[cor:#1]{Corollary~\ref*{cor:#1}}} 
\newcommand{\Alg}[1]{\hyperref[alg:#1]{Algorithm~\ref*{alg:#1}}} 
\newcommand{\Def}[1]{\hyperref[def:#1]{Definition~\ref*{def:#1}}} 

\newcommand{\Real}{{\mathbb R}}

\begin{document}

\title{Tensor-Based Fusion of EEG and FMRI to Understand Neurological Changes in Schizophrenia}

\author{\IEEEauthorblockN{Evrim Acar\IEEEauthorrefmark{1},
Yuri Levin-Schwartz\IEEEauthorrefmark{2},
Vince D. Calhoun\IEEEauthorrefmark{3}\IEEEauthorrefmark{4} and 
T\"{u}lay Adal\i\IEEEauthorrefmark{2}}
\IEEEauthorblockA{\IEEEauthorrefmark{1}Faculty of Science, University of Copenhagen, DK-1958 Frederiksberg, Denmark\\ Email: evrim@life.ku.dk}
\IEEEauthorblockA{\IEEEauthorrefmark{2} Department of Computer Science and Electrical Engineering,
University of Maryland Baltimore County, Baltimore, MD \\ Email: adali@umbc.edu, ylevins1@umbc.edu}
\IEEEauthorblockA{\IEEEauthorrefmark{4}The Mind Research Network, Albuquerque, NM, Email: vcalhoun@mrn.org}
\IEEEauthorblockA{\IEEEauthorrefmark{3} Department of Electrical and Computer Engineering, University of New Mexico, Albuquerque, NM }}

\maketitle

\begin{abstract}
Neuroimaging modalities such as functional magnetic resonance imaging (fMRI) and electroencephalography (EEG) provide information about neurological functions in complementary spatiotemporal resolutions; therefore, fusion of these modalities is expected to provide better understanding of brain activity. In this paper, we jointly analyze fMRI and multi-channel EEG signals collected during an auditory oddball task with the goal of capturing brain activity patterns that differ between patients with schizophrenia and healthy controls. Rather than selecting a single electrode or matricizing the third-order tensor that can be naturally used to represent multi-channel EEG signals, we preserve the multi-way structure of EEG data and use a coupled matrix and tensor factorization (CMTF) model to jointly analyze fMRI and EEG signals. Our analysis reveals that (i) joint analysis of EEG and fMRI using a CMTF model can capture meaningful temporal and spatial signatures of patterns that behave differently in patients and controls, and (ii) these differences and the interpretability of the associated components increase by including multiple electrodes from frontal, motor and parietal areas, but not necessarily by including all electrodes in the analysis.

\end{abstract}

\IEEEpeerreviewmaketitle

\section{Introduction}

The complexity of the human brain often necessitates the use of multiple neuroimaging techniques to better understand neural activity. Neuroimaging methods such as EEG measure the electrical activity associated with neuronal activity and provide good temporal resolution. On the other hand, methods like fMRI measure activity through changes in the blood flow and provide better spatial resolution but worse temporal resolution \cite{BuKa09}. As a result of their complementary nature, the fusion of signals from such neuroimaging methods holds much promise for better understanding the brain function.

In addition to neuroscience, in many other fields such as systems biology and recommender systems, data fusion is a topic of great interest, and joint factorization of data sets from multiple sources has proved to be a promising fusion approach \cite{2015_PIEE,lahat2015}. Data sets are coupled through either ``hard links" by extracting the same factors from the common mode or ``soft links" where the connections are established through similarity measures \cite{lahat2015,CoAdLiCa10}. If coupled data sets are in the form of matrices, they are jointly analyzed using matrix factorization-based fusion methods. In the case of coupled heterogeneous data, {\it{i.e.}}, data sets in the form of matrices and higher-order tensors, coupled matrix and tensor factorization-based methods can be used for fusion. CMTF-based approaches modeling higher-order tensors with the appropriate tensor model can capture the underlying patterns better than matrix factorization-based fusion methods \cite{2015_PIEE}.

\begin{figure}[!t]
\centering
\includegraphics[width=1.5in,trim=5 0 0 8,clip]{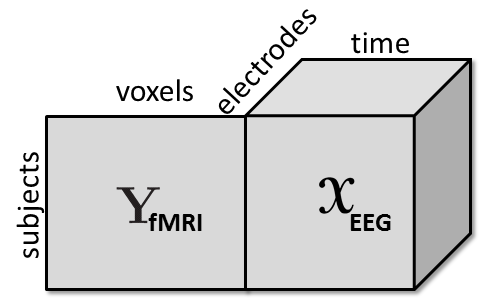}
\caption{A third-order tensor representing multi-channel EEG signals, coupled with fMRI data in the form of a matrix.}
\label{fig_EEGfMRI}
\end{figure}

Coupled heterogeneous data sets often emerge in biomedical signal processing. For instance, EEG signals from multiple subjects can be arranged as a third-order tensor with modes: \emph{subjects}, \emph{time samples} and \emph{electrodes}, and coupled with fMRI data (Figure \ref{fig_EEGfMRI}). Many studies have explored the multi-way structure of EEG signals, {\it{e.g.}}, some arranging signals from a single subject as a third-order tensor with modes: \emph{time}, \emph{frequency} and \emph{electrodes} \cite{MiMaVaNi04}, while some represent signals from several subjects as a third-order tensor as in Figure \ref{fig_EEGfMRI} \cite{Mo88}. One of the most popular tensor factorizations, {\it{i.e.}}, the CANDECOMP/PARAFAC (CP) model \cite{Ha70, CaCh70}, has proved useful in capturing patterns associated with brain activity in these studies. Nevertheless, the full multi-way structure of EEG has not been taken into account in earlier fusion studies, in particular, in joint analysis of EEG and fMRI data. Either a single electrode has been selected \cite{CaAdPeKi06} or the three-way EEG data has been unfolded as a matrix \cite{SwHuAcHuVo14}. Recently, several studies have incorporated multi-way data from various neuroimaging modalities, {\it{e.g.}}, joint factorization of EEG and magnetoencephalography (MEG) \cite{BeCoAl12}, EEG and electro-ocular artifacts (EOG) \cite{RiDuGu15}, EEG and fMRI \cite{KaRoBr15, MoSoMiGoCo04, HuPaDeHu16}. However, in these studies all electrodes are taken into account in joint analysis and it may not necessarily be better to analyze all electrodes simultaneously, in particular, if the electrodes are from different functional areas.

In this paper, we address the problem of joint analysis of fMRI and multi-channel EEG signals (represented as in Figure 1) to understand spatiotemporal differences in brain activation between patients with schizophrenia and healthy controls. Unlike EEG-fMRI studies \cite{KaRoBr15, MoSoMiGoCo04} analyzing multi-channel EEG signals of a single subject coupled with fMRI data in the \emph{spatial} or \emph{temporal} mode, we jointly analyze signals from multiple subjects by coupling the data sets in the \emph{subjects} mode. Previously, EEG and fMRI data from multiple subjects have been jointly analyzed using a CMTF model to extract signatures of interictal epileptic networks \cite{HuPaDeHu16}; however, the full potential and limitations of CMTF-based approaches for joint analysis of EEG and fMRI are not well understood. Our contributions in this paper are as follows: (i) We use subsets of electrodes from the frontal, motor and parietal areas to construct third-order tensors, and model the tensors individually using a CP model and jointly with fMRI using a structure-revealing CMTF model. (ii) We demonstrate that models can capture physically meaningful patterns that can differentiate between patients with schizophrenia and healthy controls. We discuss the advantages of the CMTF-based model and address the limitations due to modeling assumptions 
by increasing the number of electrodes included in the study. 

\section{Methodology}
In this section, we briefly describe the CP tensor factorization model and the structure-revealing CMTF data fusion model. Let the third-order tensor $\T{X} \in \Real^{I \times J \times K}$ with modes: \emph{subjects}, \emph{time samples}, and \emph{electrodes}, and matrix $\M{Y} \in \Real^{I \times M}$ (\emph{subjects} by \emph{voxels}) represent the EEG and fMRI data, respectively. An $R$-component CP model expresses tensor $\T{X}$ as a sum of third-order rank-one tensors:
\begin{small}
\begin{equation}
\T{X} \approx \sum_{r=1}^R  \MC{A}{r} \circ \MC{B}{r} \circ \MC{C}{r},
\end{equation}
\end{small}
where $\circ$ denotes the vector outer product, and $\M{A} \in \Real^{I \times R} = [\MC{a}{1} \hspace{0.02in} ... \hspace{0.02in} \MC{a}{R}], \M{B} \in \Real^{J \times R} = [\MC{b}{1} \hspace{0.02in} ... \hspace{0.02in} \MC{b}{R}],
\M{C} \in \Real^{K \times R} = [\MC{c}{1} \hspace{0.02in} ... \hspace{0.02in} \MC{c}{R}]$ correspond to factor matrices in the \emph{subjects}, \emph{time samples} and \emph{electrodes} mode, respectively. The model is unique up to permutation and scaling under certain conditions \cite{KoBa09}. If columns of factor matrices are constrained to be unit norm, the model can also be denoted as $\T{X} \approx \KOp{\V{\lambda};\M{A},\M{B},\M{C}}$, where $\V{\lambda} \in \Real^{R \times 1}$ corresponds to the weights of rank-one terms. As a result of its uniqueness leading to easily interpretable models, CP is a popular tensor factorization model. The underlying assumption by using a CP model for multi-channel EEG analysis is that each CP component models a certain brain activity pattern with certain temporal and spatial signatures, and multi-channel EEG signals from a single subject are a linear combination of those brain activities weighted by subject-specific coefficients.

When $\T{X}$ is coupled with matrix $\M{Y}$, {\it{e.g.}}, in the first mode, they can be jointly analyzed using a structure-revealing CMTF model, which has shown promise in uniquely identifying the underlying factors even in the presence of shared/unshared factors in coupled data sets \cite{AcPaGu14}. An $R$-component structure-revealing CMTF model jointly factorizes coupled data sets by minimizing the following objective function:
\begin{small}
\begin{equation} \label{eq:ACMTF}
\begin{aligned}
   f(\V{\lambda},\V{\Sigma},\M{A},\M{B},\M{C},\M{V}) & = \norm{ \T{X} - \KOp{\V{\lambda};\M{A},\M{B},\M{C}} }^2 + \norm{ \M{Y} - \smash{\M{A}\V{\Sigma}\M{V}\Tra}}^2\\
 &\hspace{10pt} + \beta \norm{\V{\lambda}}_1 + \beta \norm{\V{\sigma}}_1,
\end{aligned}
\end{equation}
\end{small}%
where the columns of factor matrices have unit norm, {\it{i.e.}},
$\norm{\MC{A}{r}} = \norm{\MC{B}{r}} = \norm{\MC{C}{r}} =
\norm{\MC{V}{r}} = 1$ for $r=1,\dots, R$. $\V{\lambda} \in \Real^{R
  \times 1}$ and $\V{\sigma} \in \Real^{R \times 1}$ correspond to the weights
of rank-one terms in $\T{X}$ and $\M{Y}$,
respectively. $\M{\Sigma} \in \Real^{R \times R}$ is a diagonal matrix
with entries of $\V{\sigma}$ on the diagonal. $\M{V} \in \Real^{M \times R}$ corresponds to the factor matrix 
in the \emph{voxels} mode. $\norm{.}$ denotes the Frobenius norm for matrices/higher-order tensors, and the 2-norm for vectors. 
$\norm{.}_1$ denotes the 1-norm of a vector, {\it{i.e.}}, $\norm{\V{X}}_1 = \sum_{r=1}^R |\VE{X}{r}|$. $\beta > 0$ is a penalty parameter. 
By imposing 1-norm penalties on the weights, (\ref{eq:ACMTF}) sparsifies the weights so that unshared factors have weights close to 0 in some data sets. Joint analysis of fMRI and EEG data using a structure-revealing CMTF model relies on the assumption that each component extracted from $\T{X}$ models a unique brain activity pattern with certain temporal and spatial signatures. The corresponding component in $\M{Y}$ (if it is a shared component) models the same brain activity and shows higher spatial specificity.
By extracting the same factor matrix $\M{A}$ from the \emph{subject} mode, in other words, using the same subject-specific coefficients to sum up the brain activity in both data sets, it is also assumed that subject covariation in fMRI and EEG are the same for shared components. 
 
\section{Experiments}
Our experiments focus on the joint analysis of fMRI and EEG data by using different subsets of electrodes to form the third-order tensors representing EEG signals. We analyze third-order tensors both individually using a CP model and jointly with fMRI data using a structure-revealing CMTF model.

\subsection{Data} 
EEG and fMRI data are separately collected from 38 subjects (22 healthy controls and 16 patients with schizophrenia) during an auditory oddball task, where subjects press a button when they detect an infrequent target sound within a series of auditory stimuli. The experimental design is described in detail in \cite{CaAdPeKi06}. For the fMRI data, we compute task-related spatial activity maps for each subject, calculated by the general linear model-based regression approach using the statistical parametric mapping (SPM) toolbox. These features form the fMRI data. For each electrode (64 electrodes in total) of the EEG data, we average small windows around the repeated instance of the target tone across the instances, deriving event-related potentials. For more details, see \cite{AdLeCa15b}.

\subsection{Experimental Setting}
Three tensors are constructed to represent EEG data using the following subsets of electrodes (i) \emph{Case 1}: Cz, Pz, Fz, (ii) \emph{Case 2}: AF3, AF4, Fz, T7, C3, Cz, C4, T8, Pz, PO3, PO4, and (iii) \emph{Case 3}: 62 electrodes excluding VEOG and HEOG. Once EEG signals are arranged as a 38 \emph{subjects} $\times$ 451 \emph{time samples} $\times$ \emph{\# of electrodes} tensor, the tensor is centered across the \emph{time} mode and scaled within the \emph{subjects} mode by dividing each horizontal slice by its standard deviation. fMRI data in the form of a 38 \emph{subjects} by 60186 \emph{voxels} matrix is preprocessed by centering each row and scaling the rows by dividing them with their standard deviations.

Each third-order tensor is modeled using a CP model using CPOPT (OPTimization) \cite{AcDuKo11a}. For fusing each tensor with fMRI data, we use ACMTF-OPT \cite{AcPaGu14} from the CMTF Toolbox for the structure-revealing CMTF model, also referred to as ACMTF (Advanced CMTF). Both models are fit using a nonlinear conjugate gradient algorithm. The penalty parameter in ACMTF is set to $\beta= 10^{-3}$. A number of random initializations are used for each model and the results returning the minimum function value are reported (making sure that models are unique). For the CP model, we use $R=3$ as the number of components for modeling all three tensors. $R>3$ results in models with lower performance or degenerate models (see \cite{KoBa09} for degeneracy). For ACMTF, we use $R=10$ components. Even for $R=10$, all components in EEG and fMRI are shared. Since both data sets measure functional characteristics of the brain, many shared components are expected.


Using an unpaired two-sample $t$-test on the columns of the factor matrix in the \emph{subject} mode, we assign a $p$-value to each component to identify the significantly differentiating components. We then assess the performance of CP and ACMTF models by the interpretability of the significant components
and their relationship to results reported in the literature.

\subsection{Results}
\begin{figure}[t!]
\centering
\begin{tabular}{c}
\subfloat[(a)]{\includegraphics[width=3.5in,trim=25 0 10 0,clip]{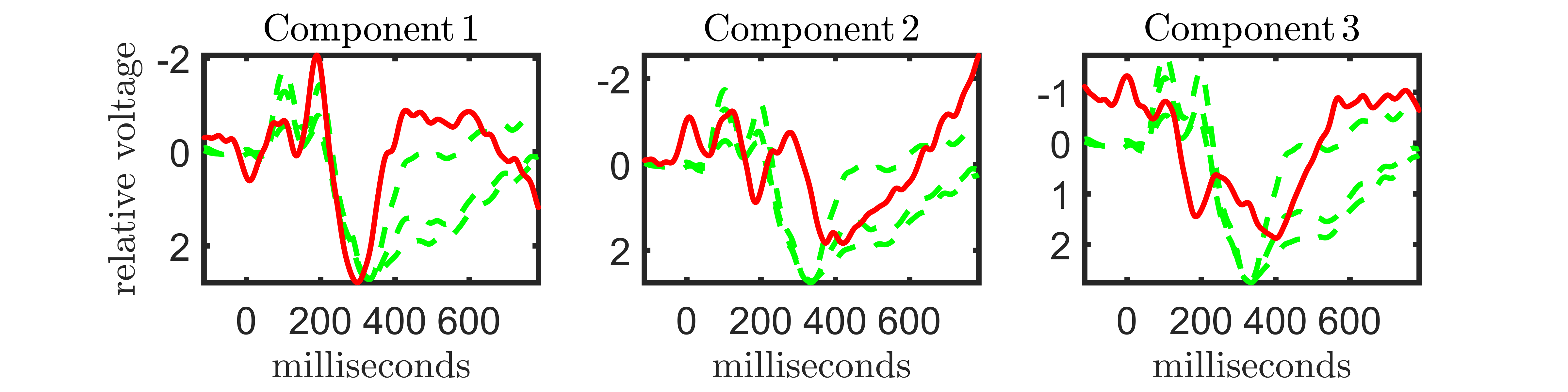}} \\
\subfloat[(b)]{\includegraphics[width=3.5in,trim=25 0 10 0,clip]{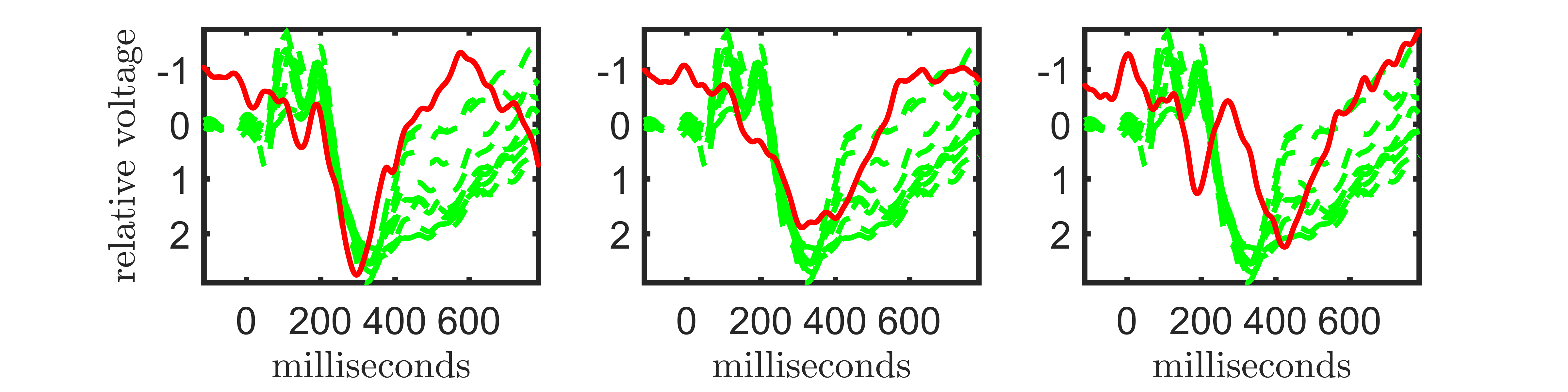}}\\
\subfloat[(c)]{\includegraphics[width=3.5in,trim=25 0 10 0,clip]{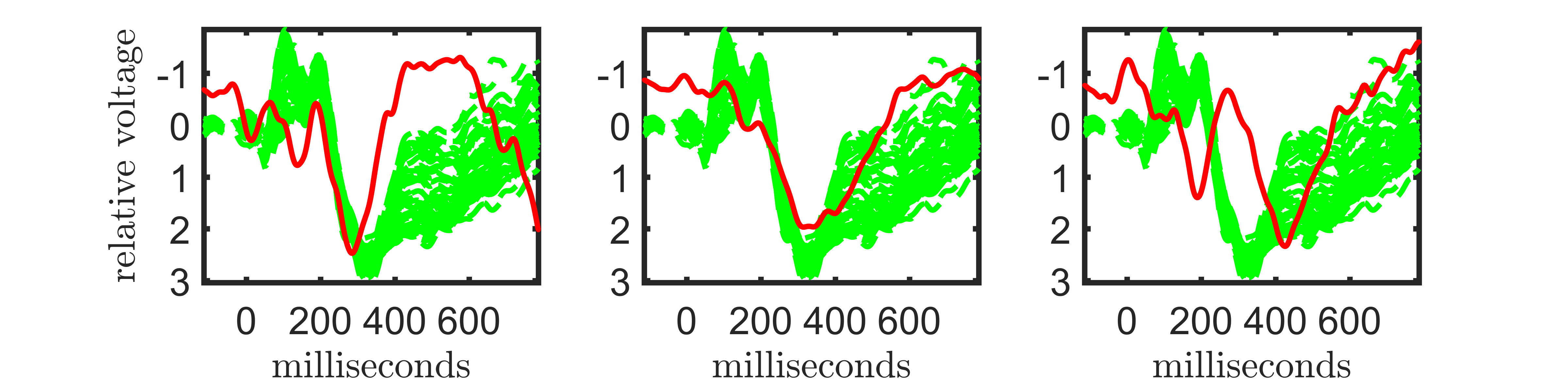}}
\end{tabular}
\caption{\small Statistically significant components for the CP models. Columns of the factor matrix in the \emph{time} mode are in red while green plots show signals from individual electrodes averaged across all subjects. The significance of components are: (a) Case 1 (3 electrodes): $5.2\times10^{-4}$, $2.2\times10^{-4}$, and 0.0098, (b) Case 2 (11 electrodes): $3.3\times10^{-5}$, $3.7\times10^{-5}$ and $6.6\times10^{-5}$, (c) Case 3 (62 electrodes): $7.7\times10^{-4}$, 0.0011 and $2.4\times10^{-4}$. Note that no correction for multiple comparisons is performed, but all components would remain significant even after the conservative Bonferroni correction.}
\label{fig:CP}
\end{figure}

CP models of the tensors constructed using different subsets of electrodes all reveal significant components. Figure \ref{fig:CP} illustrates the CP components for all cases in the \emph{time} mode. We see a close alignment of the extracted components across the three cases, which also indicates the stability of the decompositions. The first set of components 
(Component 1 for all three cases) capture the N2-P3 transition, second set (Component 2), the N2 peak, and finally the last column, the three components labeled as Component 3, capture the slow P3 response. It is worth noting that the components in Case 2 achieve the highest statistical significance values of all three cases. More importantly, as we discuss next, by fusing EEG with fMRI, we are able to link the change N2-P3 transition implicated in earlier studies \cite{CoAdLiCa10,AdLeCa15b} to the default mode network (DMN) activation in fMRI, another important biomarker for patients with schizophrenia \cite{garrity2007aberrant}.

\begin{figure*}
	\centering
\includegraphics[height=.56\linewidth]{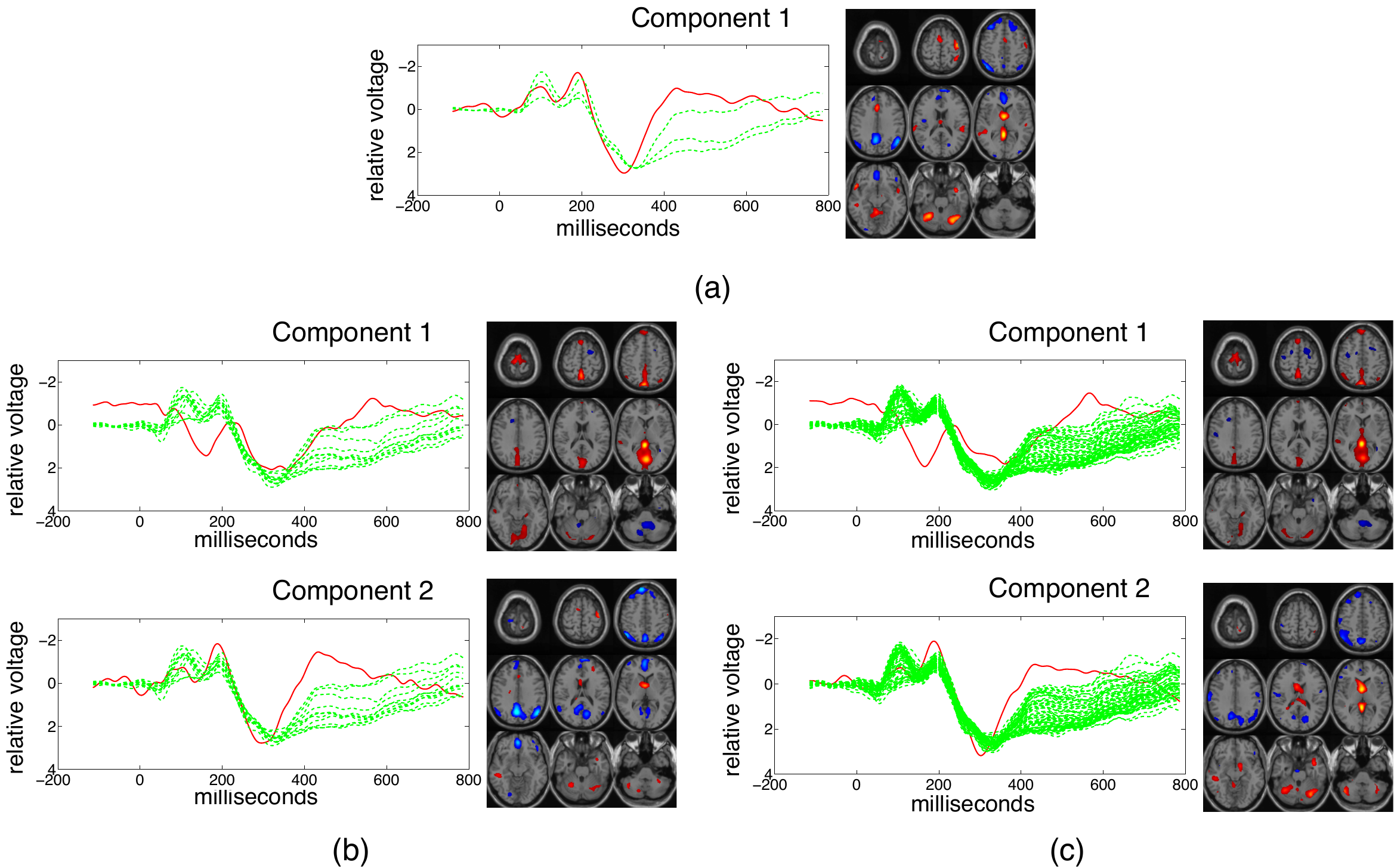}
\caption{\small Statistically significant EEG and fMRI components (surviving Bonferroni correction) derived using the ACMTF model in (a) Case 1 (3 electrodes), with significance value: 0.0011, (b) Case 2 (11 electrodes), with significance values: $3.4\times10^{-4}$ and $1.5\times10^{-4}$, and (c) Case 3 (62 electrodes), with significance values: $1.3\times10^{-4}$ and 0.0036, where the values are without Bonferroni correction. The fMRI plots are $Z$-maps, thresholded at $Z=2.7$, where red indicates an increase in controls over patients and blue indicates an increase in patients over controls. The corresponding columns of the factor matrices in the time mode (for the EEG components) are shown in red, while green plots show signals from individual electrodes averaged across all subjects.}
\label{fig:acmtf}
\end{figure*}

When the tensors are coupled with fMRI data, structure-revealing CMTF models capture several shared components with low $p$-values using 0.05 as the threshold. Figure \ref{fig:acmtf} demonstrates the components with statistically significant $p$-values that survive the Bonferroni correction in these models (only in \emph{time} and \emph{voxels} modes). For Case 1, we notice that we have a single component that shows DMN activation in fMRI, a region previously noted as affected in patients with schizophrenia during a similar auditory task \cite{garrity2007aberrant}, along with the EEG component indicating significant activation at the N2-P3 transition. For Case 2, the first component, whose EEG component describes the P3 peak, has fMRI activation in the superior parietal cortex, a region associated with logical reasoning \cite{deppe2005nonlinear}, and visual region of the brain. For the second component, EEG  corresponds to the N2-P3 transition and fMRI again to the DMN, but this time identified more robustly as indicated by the 
higher values for blue (increase in patients). In Case 3, we notice similar EEG components to the ones observed in Case 2, however the DMN is much less clear and we note noisy activation in the ventricles, similar to the component in Case 1. The N2-P3 transition component observed in the three cases is similar to the first set of aligned components in Figure 2, however now, we have tied the temporal activation with physically meaningful spatial activation in the fMRI component, namely the DMN region, thus facilitating much greater understanding of the neural disruption in schizophrenia. Note that, the significance of the components for Case 2 is higher than the component for Case 1 and around that of the components in Case 3. Thus, if interpretability and statistical significance are used as the metrics of performance, the results for Case 2 using only eleven electrodes for EEG along with fMRI are the best.

\section{Conclusion}
We have addressed the problem of jointly analyzing fMRI and EEG data from patients with schizophrenia and healthy controls, and demonstrated that CMTF-based fusion models can capture meaningful temporal and spatial signatures of patterns that can
differentiate between patients and controls. We note the gains in performance by letting the data from two modalities, EEG and fMRI fully interact. Furthermore, we have studied the effect of different subsets of electrodes and observed that the performance improves with the addition of electrodes from frontal, motor and parietal areas, but not necessarily when all electrodes are included. We plan to further study whether model order selection for the coupled model has a significant effect in these results.


\begin{thebibliography}{10}
\providecommand{\url}[1]{#1}
\csname url@samestyle\endcsname
\providecommand{\newblock}{\relax}
\providecommand{\bibinfo}[2]{#2}
\providecommand{\BIBentrySTDinterwordspacing}{\spaceskip=0pt\relax}
\providecommand{\BIBentryALTinterwordstretchfactor}{4}
\providecommand{\BIBentryALTinterwordspacing}{\spaceskip=\fontdimen2\font plus
\BIBentryALTinterwordstretchfactor\fontdimen3\font minus
  \fontdimen4\font\relax}
\providecommand{\BIBforeignlanguage}[2]{{%
\expandafter\ifx\csname l@#1\endcsname\relax
\typeout{** WARNING: IEEEtran.bst: No hyphenation pattern has been}%
\typeout{** loaded for the language `#1'. Using the pattern for}%
\typeout{** the default language instead.}%
\else
\language=\csname l@#1\endcsname
\fi
#2}}
\providecommand{\BIBdecl}{\relax}
\BIBdecl

\bibitem{BuKa09}
S.~A. Bunge and I.~Kahn, ``Cognition: An overview of neuroimaging techniques,''
  \emph{Encyc of Neuroscience}, vol.~2, pp. 1063 -- 1067, 2009.

\bibitem{2015_PIEE}
E.~Acar, R.~Bro, and A.~K. Smilde, ``Data fusion in metabolomics using coupled
  matrix and tensor factorizations,'' \emph{Proceedings of the IEEE}, vol. 103,
  pp. 1602--1620, 2015.

\bibitem{lahat2015}
D.~Lahat, T.~Adal{\i}, and C.~Jutten, ``Multimodal data fusion: A
  methodological overview methods, challenges and prospectives,''
  \emph{Proceedings of the IEEE}, vol. 103, pp. 1449--1477, 2015.

\bibitem{CoAdLiCa10}
N.~M. Correa, T.~Adal{\i}, Y.-O. Li, and V.~D. Calhoun, ``Canonical correlation
  analysis for data fusion and group inferences,'' \emph{IEEE Signal Processing
  Magazine}, vol.~27, pp. 39--50, 2010.

\bibitem{MiMaVaNi04}
F.~Miwakeichi, E.~Mart\'{i}nez-Montes, P.~A. Vald\'{e}s-Sosa, N.~Nishiyama,
  H.~Mizuhara, and Y.~Yamaguchi, ``Decomposing {EEG} data into
  space-time-frequency components using parallel factor analysis,''
  \emph{NeuroImage}, vol.~22, pp. 1035--1045, 2004.

\bibitem{Mo88}
J.~M\"ocks, ``Topographic components model for event-related potentials and
  some biophysical considerations,'' \emph{IEEE Transactions on Biomedical
  Engineering}, vol.~35, no.~6, pp. 482--484, Jun. 1988.

\bibitem{Ha70}
R.~A. Harshman, ``Foundations of the {PARAFAC} procedure: Models and conditions
  for an ``explanatory" multi-modal factor analysis,'' \emph{UCLA working
  papers in phonetics}, vol.~16, pp. 1--84, 1970.

\bibitem{CaCh70}
J.~D. Carroll and J.-J. Chang, ``Analysis of individual differences in
  multidimensional scaling via an {N}-way generalization of ``{E}ckart-{Y}oung"
  decomposition,'' \emph{Psychometrika}, vol.~35, pp. 283--319, 1970.

\bibitem{CaAdPeKi06}
V.~D. Calhoun, T.~Adal{\i}, G.~D. Pearlson, and K.~A. Kiehl, ``Neuronal
  chronometry of target detection: Fusion of hemodynamic and event-related
  potential data,'' \emph{NeuroImage}, vol.~30, pp. 544--553, 2006.

\bibitem{SwHuAcHuVo14}
W.~Swinnen, B.~Hunyadi, E.~Acar, S.~{Van Huffel}, and M.~{De Vos},
  ``Incorporating higher dimensionality in joint decomposition of {EEG} and
  {fMRI},'' in \emph{EUSIPCO}, 2014, pp. 121--125.

\bibitem{BeCoAl12}
H.~Becker, P.~Comon, and L.~Albera, ``Tensor-based processing of combined
  {EEG/MEG} data,'' in \emph{EUSIPCO}, 2012, pp. 275--279.

\bibitem{RiDuGu15}
B.~Rivet, M.~Duda, A.~Guerin-Dugue, C.~Jutten, and P.~Comon, ``Multimodal
  approach to estimate the ocular movements during {EEG} recordings: a coupled
  tensor factorization method,'' in \emph{EMBC}, 2015.

\bibitem{KaRoBr15}
E.~Karahan, P.~Rojas-Lopez, M.~Bringas-Vega, P.~Valdes-Hernandez, and P.~A.
  Vald\'{e}s-Sosa, ``Tensor analysis and fusion of multimodal brain images,''
  \emph{Proceedings of the IEEE}, vol. 103, pp. 1531--1559, 2015.

\bibitem{MoSoMiGoCo04}
E.~Martinez-Montes, P.~A. Vald\'{e}s-Sosa, F.~Miwakeichi, R.~I. Goldman, and
  M.~S. Cohen, ``Concurrent {EEG/fMRI} analysis by multiway partial least
  squares,'' \emph{NeuroImage}, vol.~22, pp. 1023--1034, 2004.

\bibitem{HuPaDeHu16}
B.~Hunyadi, W.~V. Paesschen, M.~{De Vos}, and S.~{Van Huffel}, ``Coupled
  tensor-matrix factorization of {EEG} and {fMRI} to explore epileptic network
  activity,'' in \emph{EUSIPCO}, 2016.

\bibitem{KoBa09}
T.~G. Kolda and B.~W. Bader, ``Tensor decompositions and applications,''
  \emph{SIAM Review}, vol.~51, no.~3, pp. 455--500, 2009.

\bibitem{AcPaGu14}
E.~Acar, E.~E. Papalexakis, G.~Gurdeniz, M.~A. Rasmussen, A.~J. Lawaetz,
  M.~Nilsson, and R.~Bro, ``Structure-revealing data fusion,'' \emph{BMC
  Bioinformatics}, vol.~15, p. 239, 2014.

\bibitem{AdLeCa15b}
T.~Adal{\i}, Y.~{Levin-Schwartz}, and V.~D. Calhoun, ``Multimodal data fusion
  using source separation: Application to medical imaging,'' \emph{Proceedings
  of the IEEE}, vol. 103, pp. 1494--1506, 2015.

\bibitem{AcDuKo11a}
E.~Acar, D.~M. Dunlavy, and T.~G. Kolda, ``A scalable optimization approach for
  fitting canonical tensor decompositions,'' \emph{Journal of Chemometrics},
  vol.~25, pp. 67--86, 2011.

\bibitem{garrity2007aberrant}
A.~G. Garrity, G.~D. Pearlson, K.~McKiernan, D.~Lloyd, K.~A. Kiehl, and V.~D.
  Calhoun, ``Aberrant ``default mode? functional connectivity in
  schizophrenia,'' \emph{American journal of psychiatry}, vol. 164, no.~3, pp.
  450--457, 2007.

\bibitem{deppe2005nonlinear}
M.~Deppe, W.~Schwindt, H.~Kugel, H.~Plassmann, and P.~Kenning, ``Nonlinear
  responses within the medial prefrontal cortex reveal when specific implicit
  information influences economic decision making,'' \emph{Journal of
  Neuroimaging}, vol.~15, no.~2, pp. 171--182, 2005.

\end{thebibliography}
\end{document}